\documentclass[a4paper,twoside]{article}
\usepackage{amssymb}
\usepackage{amsmath}
\usepackage{amsfonts}
\usepackage{enumerate}
\usepackage{graphicx}
\usepackage{psfrag}

\begin{document}

\title{\LARGE\bf{Background independence in a nutshell:\\ 
the dynamics of a tetrahedron}} 

\author{Daniele Colosi${}^{ab}$, Luisa Doplicher${}^{ab}$, 
Winston Fairbairn${}^{a}$,\\ Leonardo
Modesto${}^{ac}$, Karim Noui${}^{d}$, Carlo
Rovelli${}^{ab}$\\[1mm]
\em\small{${}^a$Centre de Physique Th\'eorique de Luminy,
Universit\'e de la M\'editerran\'ee, F-13288 Marseille, EU}\\[-1mm]
\em \small{${}^b$Dipartimento di Fisica dell'Universit\`a
``La Sapienza", INFN Sez.\,Roma1, I-00185 Roma, EU}\\[-1mm]
\em \small{${}^c$Dipartimento di Fisica dell'Universit\`a di Torino, INFN
Sez.\,Torino, I-10125 Torino, EU}\\[-1mm] 
\em \small{${}^d$Centre for Gravity and Geometry, Penn State 
University, University Park, PA-16802, USA.}}

\date{\small\today}

\maketitle

\begin{abstract}

We study how physical information can be extracted from a background
independent quantum system.  We use an extremely simple `minimalist'
system that models a finite region of 3d euclidean quantum spacetime
with a single equilateral tetrahedron.  We show that the physical
information can be expressed as a boundary amplitude.  We illustrate
how the notions of ``evolution" in a boundary proper-time and
``vacuum" can be extracted from the background independent dynamics.

\end{abstract}

\section{Introduction}

To understand quantum gravity, we have to learn how to do quantum
physics in a background independent context.  On a background,
distance and time separation are described by the independent
variables $x$ and $t$ that coordinatize the background.  In a
background independent field theory, on the contrary, distance and
time separation must be extracted from the dynamical variables.  In a
classical theory we know how to do so: we compare observations with
coordinate-independent quantities.  In a quantum theory, we don't.  As
a consequence, we still lack a general technique for extracting
physical information and computing, say, particle scattering
amplitudes, even when the basic formalism of a background-independent
quantum field theory is defined, as in loop quantum gravity and in the
spinfoam formalism \cite{book,ThiemannThesis,Perez}.

An idea for solving this problem is to study the quantum propagator of
a finite spacetime region, as a function of the boundary data
\cite{book,oeckl}.  The key observation \cite{cdort} is that in gravity
the boundary data include the gravitational field, hence the geometry
of the boundary, hence all relevant relative distances and time
separations.  In other words, the boundary formulation realizes very
elegantly in the quantum context the complete identification between
spacetime geometry and dynamical fields, which is Einstein's great
discovery. 

Formally, the idea consists in extracting the physical
information from a background independent quantum field theory in
terms of the quantity
\begin{equation}
W[\varphi] =\int_{\left.  \phi \right\vert_{\Sigma }=
\varphi}D\phi\  \   e^{-\frac{i}{\hbar} S[\phi]}
\ .  \label{K}
\end{equation} 
Here $\phi$ represents the ensemble of the dynamical fields, $S[\phi]$
their diffeomorphism invariant and background independent action, the
integral is over the fields in a finite spacetime region, bounded by a
compact surface $\Sigma$, and $\varphi$ is the value of $\phi$ on
$\Sigma$.  $W[\varphi]$ does not depend on (local deformations of)
$\Sigma$ because of diffeomorphism invariance.  The boundary field
$\varphi$ can be viewed as expressing initial, final as well as
boundary values of $\phi$ and $W[\varphi]$ expresses the corresponding
amplitude.  This is in the same sense in which the nonrelativistic
Feynman propagator $W(x,x',t)$ expresses the amplitude for given
initial and final positions \cite{feynhibbs}.  The difference is that
while $W(x,x',t)$ depends on the background variable $t$, here there
is no distinction between background variables, initial, final or
boundary data.  In the context of a finite number of degrees of
freedom, a covariant generalization of the Feynman propagator, viable
when there is no distinction between independent ($t$) and dependend
($x$) variable, was illustrated in \cite{Daniele}.  In field theory,
we can assume that $W[\varphi]$, formally defined in (\ref{K}),
expresses the amplitude of having a certain set of initial and final
fields, as well as boundary fields, measured by apparatus that are
located in spacetime in the manner described by (the geometry of) the
surface $\Sigma$, this geometry being determined by $\varphi$ itself. 
That is, the independent variable is simply hidden in the boundary
values of the field.  This picture closely describes what happens in a
laboratory experiment, where, say, scattering events are confined in a
finite-size spacetime region, around which we measure incoming and
outgoing particles (that is, matter-field variables) as well as
distances between instruments and elapsed time (that is,
gravitational-field variables).

We expect that particle scattering amplitudes can be effectively
computed from $W[\varphi]$ in quantum gravity; details will be given
elsewhere \cite{florian}.  The relation between particle states
defined in such a finite context and the usual particle states of
quantum field theory, defined on an infinite spacelike region, will be
discussed in \cite{daniele}.  For a theory which is not diffeomorphism
invariant, the amplitude (\ref{K}) depends also on $\Sigma$ and,
appropriately defined, can be proven \cite{cr,l} to satisfy a
generalized Tomonaga--Schwinger equation \cite{tomschw}.  This
equation becomes a generalized Wheeler-DeWitt equation in the
background independent context \cite{book}.  

This boundary picture is pithy and appealing, but its implementation
in the full 4d quantum gravity theory is difficult because of the
technical complexity of the theory.  It is useful to test and
illustrate it in a simple context.  This is what we do in this paper. 
We consider riemannian general relativity in three dimensions.  Since
the theory is topological, the integral (\ref{K}) is trivial.  To
further simplify the context, we triangulate spacetime, reducing the
field variables to a finite number \cite{Regge,ooguri,makela}. 
Furthermore, we take a `minimalist' triangulation: a single
tetrahedron with four equal edges.  In this way the number of
variables we deal with is reduced to a bare minimum.  The result is an
extremely simple system, which, nevertheless, is sufficient to realize
the conceptual complexity of a background independent theory of
spacetime geometry.

We show that this simple system has in fact a background independent
classical and quantum dynamics.  The classical dynamics is governed by
the relativistic Hamilton function \cite{book}, the quantum dynamics
is governed by the relativistic propagator (\ref{K}).  We compute both
these functions explicitly.  The classical dynamics, which is
equivalent to the Einstein equations, fixes relations between
quantities that can be measured on the boundary of the tetrahedron. 
The quantum dynamics gives probability amplitudes for ensembles of
boundary measurements.

The model and its interpretation are well-defined with no need of
picking a particular variable as a time variable.  However, we can
also identify an elapsed proper ``time" $T$ among the boundary
variables, and reinterpret the background independent theory as a
theory describing evolution in the observable time $T$ (observables in
the sense of \cite{Partial}, see \cite{book}.)  We describe the two
(equivalent) interpretations of the model, in the classical as well in
the quantum theory.  Furthermore, we concretely illustrate the
distinction between the nonperturbative vacuum state and the
``Minkowski" vacuum that minimizes the energy associated with the
evolution in $T$, and we show that the technique suggested in
\cite{cdort} for computing the Minkowski vacuum state from the
nonperturbative vacuum state works in this context.

Thus, the system captures the essence of background independent
physics in a nutshell.

The classical theory is discussed in Section 2; classical time
evolution in Section 3; the quantum theory in Section 4; quantum time
evolution in Section 5.  As a preliminary step, we describe below the
geometry of an equilateral tetrahedron.

\subsection{Elementary geometry of an equilateral tetrahedron}

\psfrag{a}{$a$}
\psfrag{b}{$b$}
\psfrag{c}{$c$}

\begin{figure}
  \begin{center}
  \includegraphics[height=3cm]{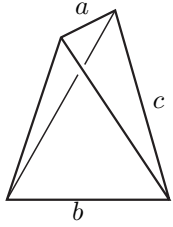}
  \end{center}
  \caption{\label{equi} The equilateral tetrahedron.}
  \end{figure}

Consider a tetrahedron immersed in euclidean three-dimensional space. 
Let $a$ be the length of one of the edges (we call it the ``top" edge)
and $b$ the length of the opposite (``bottom") edge, namely the edge
disjoint from the top edge.  Assume that the other four (``side")
edges have equal length $c$.  See Figure \ref{equi}.  We call such a
tetrahedron ``equilateral".  We call
$\theta_{a}, \theta_{b}, \theta_{c},$ the (respectively ``bottom",
``top" and ``side") dihedral angles at the edges (with length) $a$,
$b$, $c$.  Elementary geometry gives
\begin{equation}
\sin{\frac{\theta_{a}}{2}}=\frac{b}{\sqrt{4c^2-a^2}},\hspace{3em} 
\sin\frac{\theta_{b}}{2}=\frac{a}{\sqrt{4c^2-b^2}},\hspace{3em} 
\cos\theta_c = \frac{ab}{\sqrt{(4c^2-a^2)(4c^2-b^2)}}. 
\label{angles}
\end{equation}
(The last equation can be easily obtained from the scalar product of
the normals to two adjacent triangles, by working in the orthonormal
basis determined by the top and bottom edges, and the tetrahedron
axis.)  It follows from (\ref{angles}) that
\begin{equation}
\cos\theta_c = \sin{\frac{\theta_{a}}{2}}\ \sin{\frac{\theta_{b}}{2}}.
\label{key}
\end{equation}

For later purpose, we consider also the case in which $c \gg a,b$.  In 
this case, we have, to the first relevant order,
\begin{equation}
\theta_{a}=\frac{b}{c},\hspace{3em} 
\theta_{b}=\frac{a}{c},\hspace{3em} 
\theta_c = \frac{\pi}{2}-\frac{ab}{4c^2}
\label{anglesl}
\end{equation}
and
\begin{equation}
\theta_c  = \frac{\pi}{2}-\frac{\theta_{a}\theta_{b}}{4}. 
\label{keyl}
\end{equation}

We consider also the three external angles at the edges
\begin{equation}
k_{a}(a,b,c)=\pi-\theta_{a}(a,b,c), \hspace{2em}
k_{b}(a,b,c)=\pi-\theta_b(a,b,c)  ,\hspace{2em}
k_{c}(a,b,c)=\pi-\theta_{c}(a,b,c). 
\label{phi}
\end{equation}
Notice that they express the discretized extrinsic curvature of the
surface of the tetrahedron.  This is why we have denoted them with the
letter $k$, often used for the extrinsic curvature.  Using
(\ref{angles}) and (\ref{phi}), the relation between the edge lengths
$a,b,c$ and the external angles ${k}_{a},{k}_{b},{k}_{c}$ can be
written in the form
\begin{eqnarray}
a &=& \sqrt{4c^2-b^2}\ \cos \frac{{k}_{b}}{2},\nonumber \\
b &=& \sqrt{4c^2-a^2}\ \cos  \frac{{k}_{a}}{2}, \nonumber \\
ab &=& -\sqrt{(4c^2-a^2)(4c^2-b^2)}\  \cos {k}_{c}\ ;
\label{em3}
\end{eqnarray}
while (\ref{key}) reads 
\begin{equation}
\cos{k}_c = -\cos{\frac{{k}_{a}}{2}}\ \cos{\frac{{k}_{b}}{2}}.
\label{keyphi}
\end{equation}

\section{Classical theory}

\subsection{Regge action}

Consider the action of general relativity, in the case of a simply
connected finite spacetime region $\cal R$.  Recall that in the
presence of a boundary $\Sigma=\partial{\cal R}$ we have to add a
boundary term to the Einstein--Hilbert action, in order to have well
defined equations of motion.  The full action reads
\begin{equation}
S_{\rm GR}[g] = \int_{\cal R} d^nx \sqrt{\det{g}}\ R + \int_{\Sigma}
d^{n-1}x \sqrt{\det{q}}\ k.
\label{actionGR}
\end{equation}
Here $g$ is the metric field, $R$ is the Ricci scalar, $n$ is the
number of spacetime dimensions, while $q$ is the metric, and $k$ the
trace of the extrinsic curvature, induced by $g$ on $\Sigma$.  (For a
discussion on different choices of boundary terms in three-dimensional
gravity, see \cite{loug}; here we are interested in variations of the
action at fixed value of the boundary metric.)

In general, the \emph{Hamilton function} of a finite dimensional
dynamical system is the value of the action of a solution of the
equations of motion, viewed as a function of the initial and final
coordinates; the general solution of the equations of motion can be
obtained from the Hamilton function by simple derivations
\cite{hamilton}.  In field theory, the Hamilton function can be
defined as the value of the action of a solution of the equations of
motion, integrated on a finite region $\cal R$, viewed as a function
of value of the field on the boundary $\Sigma$ (see \cite{book}).  In
general relativity, the Hamilton function $S[q]$ is the value of the
action (\ref{actionGR}), computed on the solution $g_{q}$ of the
equations of motion determined by the boundary value $q$
\begin{equation}
S[q] = S_{\rm GR}[g_{q}]. 
\label{Hamilton}
\end{equation}
If $g_{q}$ is not unique, $S[q]$ is multivalued.  Notice that $S[q]$
is independent from (local deformations of) $\Sigma$, because of
diffeomorphism invariance.

Since the bulk action vanishes on a vacuum solution of the equations
of motion, the Hamilton function of general relativity reads
\begin{equation}
S[q] = \int_{\Sigma} d^{n-1}x\, \sqrt{\det{q}}\ k[q]. 
\label{HGR}
\end{equation}
where the extrinsic curvature $k[q]$ is a nonlocal function,
determined by the Ricci-flat metric $g_{q}$ bounded by $q$.

In the following we consider only the three-dimensional riemannian
case, where $n=3$ and the signature of $g$ is $[+++]$.  In this case,
we must add an overall minus sign in (\ref{actionGR}) and (\ref{HGR}),
see for instance the Appendix C of reference \cite{Hawk}. 
Furthermore, we consider the discretization of the theory provided by
a Regge triangulation \cite{Regge}.  Let $i$ be the index labelling
the links of the triangulation and call $l_{i}$ the length of the link
$i$.  In three dimensions, the bulk Regge action is
\begin{equation}
S_{\rm Regge}(l_{i}) = -\sum_{i}\ l_{i}\left(2\pi-\sum_{t} \ 
\theta_{i,t}(l)\right), 
\label{Regge}
\end{equation}
where $\theta_{i,t}(l)$ is the dihedral angle of the tetrahedron $t$
at the link $i$, and the angle in the parenthesis is therefore the
deficit angle at $i$.  The boundary term is
\begin{equation}
S_{\rm boundary}(l_{i}) = -\sum_{{\rm boundary}\ i} l_{i}
\left(\pi -\sum_{t} \theta_{i,t}(l)\right).
\label{Reggeb}
\end{equation}
Notice that the angle in the parenthesis is the angle formed by the
boundary, which can be seen as a discretization of the extrinsic
curvature.

We choose the minimalist triangulation formed by a single
tetrahedron, and, furthermore, consider only the case in which the
tetrahedron is equilateral.  Then there are no internal links, the
Regge action is the same as the Regge Hamilton function, and is given
by
\begin{equation}
S(a,b,c) = -a\ {k}_{a}(a,b,c) -b\ {k}_{b}(a,b,c) -4c \ 
{k}_{c}(a,b,c). 
\label{action1}
\end{equation}
The expression for the dihedral angles as functions of the edges
length, for a flat interior geometry, is given in (\ref{angles}) and
(\ref{phi}).  Inserting these equations into (\ref{action1}) gives
the Hamilton function
\begin{eqnarray}
S(a,b,c) &=&  a  \, \left(2 
\arcsin{\frac{b}{\sqrt{4c^2-a^2}}}-\pi\right)
          + b \, \left(2 
\arcsin{\frac{a}{\sqrt{4c^2-b^2}}}  -\pi\right) \nonumber \\
          && +\  4c \, \left( 
\arccos{\frac{ab}{\sqrt{4c^2-a^2}\; 
\sqrt{4c^2-b^2}}} -\pi \right).
\label{action}
\end{eqnarray}

\subsection{The dynamical model and its physical meaning}

The Hamilton function (\ref{action}) defines a simple relativistic
dynamical model.  The model has three variables, $a,b$ and $c$.  These
are partial observables in the sense of \cite{Partial}.  That is, they
include both the independent (``time") and the dependent (dynamical)
variables, all treated on equal footing.  (This paper is self
contained, but the general formalism and the interpretation of these
general relativistic systems is discussed in detail in \cite{book}.)

The equations of motion are obtained following the general algorithm
of the relativistic Hamilton--Jacobi theory \cite{book}: define the
momenta
\begin{equation}
p_{a}(a,b,c) = \frac{\partial S(a,b,c)}{\partial a}, \hspace{2em}
p_{b}(a,b,c) = \frac{\partial S(a,b,c)}{\partial b}, \hspace{2em}
p_{c}(a,b,c) = \frac{\partial S(a,b,c)}{\partial c},
\end{equation}
and equate them to constants 
\begin{equation}
p_{a}(a,b,c) = p_{a}, \hspace{2em}
p_{b}(a,b,c) = p_{b}, \hspace{2em}
p_{c}(a,b,c) = p_{c}. 
\label{em}
\end{equation}
These equations give the dynamics, namely the solution of the
equations of motion.  Explicitly, the calculation of the momenta is
simplified by the observation that the action is a homogeneous
function of degree one, hence
\begin{equation}
S(a,b,c) = a\ \frac{\partial S(a,b,c)}{\partial a}
+b\ \frac{\partial S(a,b,c)}{\partial b}
+c\ \frac{\partial S(a,b,c)}{\partial c};
\end{equation}
this allows us to identify immediately 
\begin{equation}
p_{a}(a,b,c) =  -{k}_{a}(a,b,c), \hspace{2em} p_{b}(a,b,c) = 
 -{k}_{b}(a,b,c), \hspace{2em} p_{c}(a,b,c)  = 
      -4\ {k}_{c}(a,b,c).
\label{em2}
\end{equation}
Inserting the explicit form (\ref{angles}) of the angles, 
we obtain the evolution equations
\begin{eqnarray}
a &=& \sqrt{4c^2-b^2}\ \cos \frac{p_{b}}{2},\nonumber \\
b &=& \sqrt{4c^2-a^2}\ \cos  \frac{p_{a}}{2}, \nonumber \\
ab &=& -\sqrt{(4c^2-a^2)(4c^2-b^2)}\  \cos \frac{p_{c}}{4},
\label{em33}
\end{eqnarray}
which reproduce (\ref{em3}).
This result deserves various comments.  
\begin{enumerate}[(i)]\addtolength{\itemsep}{-.3mm}
\item We begin with a technical comment.  Notice that the variation of
the action with respect to the lengths is completely determined by the
variation of the first length factor in (\ref{Reggeb}): the variation
of the length in the argument of the angles has no effect on the
action.  The fact that this variation vanishes was already pointed out
by Regge \cite{Regge}.  It is the discrete analog of the well-known
fact that in deriving the Einstein equations from the
Einstein--Hilbert action we can ignore the change of the Levi--Civita
connection under a variation of the metric. 

\item Notice that boundary lengths $a$, $b$, $c$ determine the
intrinsic geometry of the boundary surface.  Their conjugate momenta
$p_{a}$, $p_{b}$, $p_{c}$, are determined by the dihedral angles and
are given by the external angles at the links.  That is, they measure
the extrinsic curvature of the boundary surface.  This is precisely as
in the ADM hamiltonian framework \cite{ADM}, where the momentum
variable conjugate to the metric is the extrinsic curvature.  Equation
(\ref{em2}) is the discrete analog of the ADM relation between
momenta and extrinsic curvature.

\item The evolution equations (\ref{em33}) are not independent, as is
always the case in relativistic systems (for instance, out of the four
equations of motion of a relativistic particle, only three are
independent).  We can take the first two equations as the
independent ones.  They express relations between the lengths and
dihedral angles of the tetrahedron.
\item 
How are the evolution equations (\ref{em33}) related to the Einstein
equations?  They are essentially equivalent.  In three
dimensions, the vacuum Einstein equations $R_{\mu\nu}=0$, where
$R_{\mu\nu}$ is the Ricci tensor, imply that the Riemann tensor
vanishes, namely that spacetime is flat.  This implies that the
tetrahedron is immersed in a {\em flat\ } 3d spacetime.  But if
spacetime is flat, the extrinsic curvature of the boundary at the edge
is exactly equal to $\pi$ minus the dihedral angle.  Hence these
equations express the flatness of spacetime, namely they have the same
content as the Einstein equations $R_{\mu\nu}=0$.  In other words, we
have derived the relation (\ref{angles}) between length and angles
assuming a {\em flat} 3d space: viceversa, the fact that these
relations are satisfied implies that, in the approximation captured by
the triangulation, 3d space is flat, namely the Einstein equations
hold.
\item
The physical interpretation of the model is as follows.  We assume
that we can measure the three lengths $a$, $b$ and $c$ and the three
external angles ${k}_{a}$, ${k}_{b}$ and ${k}_{c}$ (these are six
partial observables in the sense of \cite{Partial}).  These are all
local observations that can be made on the boundary surface.  They
refer to the intrinsic as well as the extrinsic geometry of the
surface itself.  The classical theory establishes relations between
these measurable quantities.  These relations are the physical content
of the theory and are given by the equations (\ref{em33}).  They are
equivalent to the statement that spacetime is flat (to the given 
approximation).
\item
The fact that the equations of motion are not independent
is reflected in a relation between the momenta.  The relation is of
course the one given by equation (\ref{keyphi}), that is
\begin{equation}
H(p_{a},p_{b},p_{c}) = \cos{\frac{p_{c}}{4}} + \cos{\frac{p_{a}}{2}}\  
\cos{\frac{p_{b}}{2}} = 0. 
\label{key2}
\end{equation}
{} From this we can directly read out the Hamilton--Jacobi equation
satisfied by $S(a,b,c)$.
\begin{equation}
\cos \frac{1}{4}\frac{\partial S}{\partial c} 
+ \cos{\frac{1}{2}\frac{\partial S}{\partial a}}\  
\cos{\frac{1}{2}\frac{\partial S}{\partial b}} \ =\ 0 . 
\label{HJ}
\end{equation}
The function $H(p_{a},p_{b},p_{c})$ given in (\ref{key2}) is the
\emph{relativistic hamiltonian} \cite{book}, or hamiltonian
constraint, of the system.

\item Finally, in the limit in which $c \gg a,b$, the action is given
simply by
\begin{equation}
S(a,b,c) =  \frac{ab}{c}- (a+b+2c)\,\pi,
\label{largeT}
\end{equation}
and the evolution equations (\ref{em33}) become 
\begin{eqnarray}
a = c\ (p_{b}+\pi), 
\ \ \ \ \ \ \ \ 
b = c\ (p_{a}+\pi), 
\ \ \ \ \ \ \ \ 
ab = -c^2\ (p_c+2\pi).
\label{em333}
\end{eqnarray}

\end{enumerate}

\section{Time evolution}

In the description given so far, no reference to evolution in a
preferred time variable was considered.  We now introduce it here.  We
decide to regard the direction of the axis of the equilateral
tetrahedron as a temporal direction.  In particular, we decide to
interpret $b$ as an initial variable and $a$ as a final variable ($b$
for \emph{before} and $a$ for \emph{after}).  The length $c$ of the
side links can then be regarded as a (proper!)  length measured in the
temporal direction, namely as the physical time elapsed from the
measurement of $a$ to the measurement of $b$.  Indeed, had we
considered a spacetime with signature $[++-]$, and assuming we had
oriented the tetrahedron axis in a timelike direction, $c$ would {\em
precisely} be the physical time measured by a real clock on the
boundary of the spatial region considered, the worldline of the clock
running along one of the side edges.

To emphasize this interpretation of the variable $c$, in this section
we change its name, renaming $c$ as $T$.  The Hamilton function reads
then $S(a,b,T)$ and can now be interpreted as the Hamilton function
that determines the evolution in $T$ of a variable $a$.  The variable
$b$ is interpreted as measured at time $T=0$ and the variable $a$ at
time $T$; therefore $b$ can be viewed as an integration constant for
the evolution of $a$ in $T$.  (Notice that $b$ is not necessarily the
same variable as $a$, namely $T=0$ does not imply $a=b$).

For comparison, recall that the Hamilton function of a free particle
moving from a position $b$ to a position $a$ in a time $T$ is
\begin{equation}
S_{\rm free\  particle}(a,b,T) = \frac{m(a-b)^2}{2T}
\label{freeparticele}
\end{equation}
which completely describes the free particle dynamics: equations
(\ref{em}) give in fact 
\begin{eqnarray}
p_{a}(a,b,T) &=& \frac{\partial S(a,b,T)}{\partial a} = m\frac{a-b}{T} = 
p_{a}, \\ 
p_{b}(a,b,T) &=& \frac{\partial S(a,b,b)}{\partial a} = m\frac{b-a}{T}
= p_{b}, \\
p_{c}(a,b,T) &=& \frac{\partial S(a,b,T)}{\partial T} = -
\frac{m(a-b)^2}{2T^2} =  p_{T},
\end{eqnarray}
which can be readily recognized as the evolution equation for 
coordinate and momentum
\begin{equation}
a(T) = a_{0} + V T, \hspace{3em} p_{a}(T) = m V,
\end{equation}
where $a_{0}=b$ and $V=-p_{b}/m$, and the relation between energy
($E\equiv-p_{T}$) and momentum
\begin{equation}
E=H(p_{b})=\frac{p_{b}^2}{2m}, 
\end{equation}
which defines the hamiltonian function $H(p_{b})$.

Returning to our system, the hamiltonian that evolves the system in
the time $T$, which we can call ``proper-time hamiltonian", can
obtained from the energy
\begin{equation}
E= - p_{T}= -\frac{\partial S(a,b,T)}{\partial T}  =
4 \, \pi -
 4 \, 
\arccos\frac{ab}{\sqrt{(4T^2-a^2)(4T^2-b^2)}} 
\label{energyPi}
\end{equation} 
by using the equations of motion to express the initial position $b$
as a function of the position $a$ and momentum $p_{a}$.  This gives
\begin{eqnarray}
H(a,p_{a},T) & = & 4 \, 
\pi - 4\arccos\left(\frac{a\cos(p_a/2)}{\sqrt{4 T^2 - (4 
T^2 
- a^2) \cos^2(p_a/2)}} \, 
 \right) 
\label{hamiltonsPi}
\end{eqnarray}
Notice that the angle $\theta_{c}$ can vary between $0$ and $\pi/2$,
and therefore so does the $\arccos $.  Therefore the energy can vary
between $2\pi$ and $4\pi$.  The fact that the domain of the energy is
bounded has important consequences.  For instance, we should expect
time to become discrete in the quantum theory.

In this way, the relativistic background independent system can be
reinterpreted as an evolution system, where the ``proper time" on the
boundary of the region of interest is taken as the independent time
variable.  The Hamilton equation generated by the hamiltonian for
$a(T)$ and $p_a(T)$ are:
\begin{eqnarray}
\frac{d a(T)}{dT} = \frac{\partial H}{\partial p_{a}} 
&=& \frac{4 a T}{4 T^2 \sin^2(p_a/2) + a^2 \cos^2(p_a/2)} ,
\nonumber \\
\frac{d p_a(T)}{dT} = - \frac{\partial H}{\partial a} &=& -\frac{4 T
\sin(p_a)}{4 T^2 \sin^2(p_a/2) + a^2\cos(p_a/2)}.
\label{eq.esatte}
\end{eqnarray} 
The solution of these equations is
\begin{eqnarray}
a(T) & = & \sqrt{4T^2-b^2}\ \cos \frac{p_{b}}{2}, \nonumber \\
p_{a}(T) & = &-2\arccos\frac{b}{\sqrt{4T^2
\sin^2(p_{b}/2)+b^2\cos^2(p_{b}/2)}},
\label{em4}
 \label{impli.eq}
\end{eqnarray}
where $b$ and $p_{b}$ are integration constants.  These solutions are
immediately recognized as the equations (\ref{em3}).  Therefore the
dynamics generated by the hamiltonian is the same as the general
relativistic dynamics defined in a-temporal terms in the previous
section.

\psfrag{T}{$T$}

\begin{figure}
  \begin{center}
  \includegraphics[height=5cm]{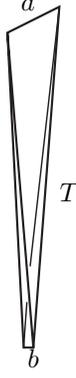}
  \end{center}
  \caption{\label{lungo} For large $T$, at constant
  $b$ and $\theta_{b}$, we have $\theta_{a}\to 0$ and $a\sim T$.}
  \end{figure}

It is interesting to consider the long time evolution of the system. 
In the large $T$ limit we have the behavior
\begin{equation}
a(T) \to {\rm const}\ T, \hspace{2em}
p_{a}(T) \to \frac{\rm const}{T} - \pi,
\label{largeTsol}
\end{equation}
which is precisely (\ref{anglesl}), identifying the two integration
constants with the initial data $\theta_{a}$ and $a$.  Therefore
$p_{a}(T)$ tends to $-\pi$ as $T$ increases.  It is easy to understand
this behavior geometrically. See Figure \ref{lungo}: at fixed values
of the bottom length $b$ and bottom angle $\theta_{b}=\pi+p_{b}$, as
the side length $T$ grows, we have that the top angle
$\theta_{a}=\pi+p_{a}\to 0$ and $a$ grows proportionally to $T$.

The energy is not constant (there is no reason for the energy to be
constant) and tends to
\begin{equation}
E(T) \to 2\pi 
\label{largeTen}
\end{equation}
which is its minimal value.  This result can also be obtained by
considering the hamiltonian for large $T$.  Starting from
(\ref{largeT}), we obtain
\begin{equation}
H = -\frac{\partial S(a,b,T)}{\partial T} = \frac{ab}{T^2}
+ 2\pi =  \frac{a(\pi-p_{a})}{T} + 2\pi .
\end{equation}
The equations of motion 
\begin{eqnarray}
\frac{d a(T)}{dT} = \frac{\partial H}{\partial p_{a}} 
= \frac{a}{T} ,
&\hspace{3em}&
\frac{d p_a(T)}{dT} = - \frac{\partial H}{\partial a} = -\frac{
\pi + p_{a}}{T},
\label{ap}
\end{eqnarray} 
are solved by (\ref{largeTsol}) and yield (\ref{largeTen}).  

Notice that the convergence of the ``velocity" to the attraction point
$p_{a}\to-\pi$ and the energy to its minimal value, resembles a
dissipative system, such as a point particle under a constant force
in a fluid. 

\subsection{Phase space and extremal configurations}

\begin{figure}
  \begin{center}
  \includegraphics[height=1cm]{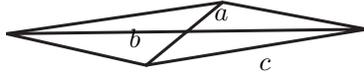}
  \end{center}
  \caption{\label{flat} The flat tetrahedron: the bottom and top edges
  touch.}
  \end{figure}

Viewed as a dynamical system evolving in $T$, our system has a phase
space coordinatized by $a\in [0,\infty[$ and $p_{a}\in [0,-\pi]$.  The
maximum value of the energy (\ref{hamiltonsPi}) on this phase space is
$E_{min} = 4\pi$, which is attained along the boundary $p_{a}=0$ of
$\Gamma$.  These are states with vanishing external angle at the top
edge.  They are configuration in which the tetrahedron is
``flattened": its volume is zero, and the upper and bottom edges
touch.  The value of $a$ is arbitrary.  See Figure \ref{flat}.  Notice
that these configurations evolve into one another.  In fact, if
$p_{a}=0$, (\ref{em4}) gives
\begin{eqnarray}
a(T) &=& \sqrt{4T^2-b^2},\nonumber \\
p_{a}(T) &=&0.
\label{em4vacuum}
\end{eqnarray}
Therefore these states grow in $T$ remaining flattened and with the
the energy remaining constant in $T$ at the value $E=4\pi$.  

In all the other states, the energy changes with time.  As $T$ grows a
generic state evolves towards a state of the form
\begin{eqnarray}
a &=& 2T\ \cos \frac{p_{b}}{2},\nonumber \\
p_{a} &=& -\pi,
\label{attractive}
\end{eqnarray}
with the energy converging to the value $E=2\pi$.  These states 
minimize the energy and form the boundary $p_{a}=-\pi$ of $\Gamma$.  We
call these states ``Minkowski vacuum states", since they minimize
the energy.  Notice that their definition depends on the choice of the
time variables made.

\begin{figure}[ht]
  \begin{center}
  \includegraphics[height=6cm]{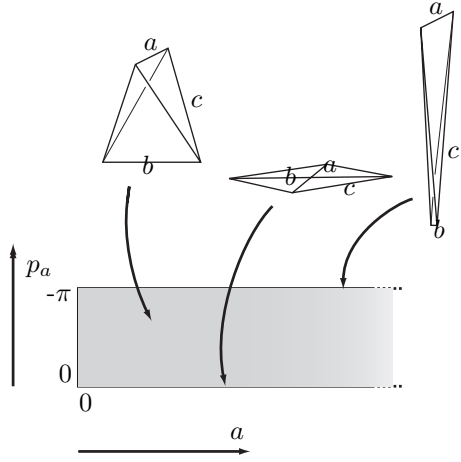}
  \end{center}
  \caption{\label{fasi} The phase space of the system with some
  typical configurations.  The ``Minkowski" states are the ones along
  the $p_{a}=-\pi$ boundary.}
  \end{figure}

Therefore the 2d phase space has two notable subsets: the line
$p_{a}=0$ forms an independent sector evolving into itself, given by
the energy-maximizing states; while the line $p_{a}=-\pi$ is an
attractor for the rest of the phase space, and is formed by the
energy-minimizing states that we have called ``Minkowski" states.  See
Figure \ref{fasi}

\psfrag{p}{$p_{a}$}
\psfrag{r}{-$\pi$}
\psfrag{0}{$0$}

Notice that the variable $T$ is bounded by $|T|>b/2$ from (\ref{em4}),
therefore we cannot continue the solution for arbitrarily small $T$. 
It is natural to introduce the time variable
\begin{eqnarray}
t&=&\sqrt{T^2-b^2/4},\ {\rm for}\ T>b/2 
\end{eqnarray}
which geometrically represents the height of the triangular face of
the tetrahedron with base $b$, and which arrives at zero.  The
evolution equations read then
 \begin{eqnarray}
a(T) & = & 2t \cos \frac{p_{b}}{2}, \nonumber \\
p_{a}(T) & = &-2\arccos\frac{b}{\sqrt{4t^2
\sin^2(p_{b}/2)+b^2}}.
\label{em5}
\end{eqnarray}

\begin{figure}[ht]
  \begin{center}
  \includegraphics[height=6cm]{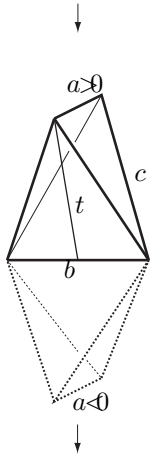}
  \end{center}
  \caption{\label{glove} The extension to negative $a$ and negative 
  $t$.}
  \end{figure}

Notice that the equations of motions can be extended also for
\emph{negative} $t$ and \emph{negative} $a$ and $p_a$.  It is natural
to interpret this as an evolution in which the tetrahedron crosses
the point $a=0, p_{a}=0$ in which it has zero volume, and grows ``on
the other side", overturned as a glove.  See Figure \ref{glove}. If 
we consider this extension, we can take the phase space to be given by 
$a\in R$ and $p_{a}\in [-\pi,\pi]$.   In the following, we will not 
consider this extension.

\psfrag{t}{$t$}
\psfrag{<}{$<$}
\psfrag{>}{$>$}

\section{Quantum theory}

\subsection{Kinematics}

We begin by constructing the boundary Hilbert space $\cal K$, on which
the operators representing the boundary (partial) observables are
defined.  Consider the triad formalism for three-dimensional euclidean
general relativity.  The variables are a the triad $e_\mu^i(x),\
\mu=1,2,3,\ i=1,2,3$ and its $SO(3)$ spin connection $A_{\mu}^i(x)$. 
The canonical boundary variables can be taken to be $SO(3)$ connection
$A_{a}^i(x)\ a= 1,2$ and the inverse densitized triad $E^a_{i}(x)$
induced on the boundary surface.  Let us then discretize spacetime in
terms of a single tetrahedron $T$.

Call $f^{p},\ p=1,2,3,4$, the faces of the tetrahedron, $e^{pq}$ the
oriented edge separating the face $p$ from the face $q$ (say oriented
rightward in going from $p$ to $q$).  To define the discrete dynamical
variables, consider the dual tetrahedron $T^*$ defined by vertices
$v_{p}$ in the face $f^{p}$ of $T$.  The edges $e_{pq}$ of $T^*$
connect the vertex $p$ to the vertex $q$; they are dual, and cut the
corresponding edges $e^{pq}$ of $T$.  We can discretize the boundary
field $A_{a}^i(x)$ by replacing it with six group elements $U_{pq}$
associated to the six edges $e_{pq}$, interpreted as the parallel
transport matrix of the connection along $e_{pq}$.  As usual in
quantum gravity, we take $U_{pq}\in SU(2)$ (the classical theory is
determined by the algebra, not the group).  We write
$U_{pq}=U_{qp}^{-1}$.  Gauge transformations act on the vertices
$v_{p}$; they are determined by four group elements $V_{p}$ and the
group elements $U_{pq}$ transform as
\begin{equation}
U_{pq}\ \to \ V_{p} U_{pq} V_{q}^{-1}. 
\label{gaugeU}
\end{equation}

\begin{figure}
  \begin{center}
  \includegraphics[height=3cm]{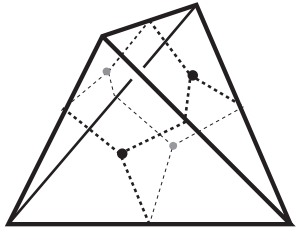}
  \end{center}
  \caption{\label{dual} The tetrahedrons $T$ (continuous lines)
  and $T^*$ (pointed lines).}
  \end{figure}

The quantum theory can be defined starting from the Hilbert space
$\cal K$ of the Haar-square-integrable functions $\psi(U_{pq})$ of the
six dynamical variables $U_{pq}$ that are gauge-invariant under the
transformations (\ref{gaugeU}), namely
\begin{equation}
\psi(U_{pq}) = \psi(V_{p} U_{pq} V_{q}^{-1}).
\end{equation}
These gauge transformations depend on four group elements, therefore
${\cal K}={\cal L}_2[(SU(2))^6/(SU(2))^4]$ where the action of
$(SU(2))^4$ on $(SU(2))^6$ is the one given in (\ref{gaugeU}). 
We use the notation ${\mathbf U} = (U_{pq})$ for the 6-tuplet of group
elements, and thus write states as $\psi({\mathbf U})$.  Similarly, 
we indicate an 6-tuplet of spins as ${\mathbf j}=(j_{pq})$. As
well known (see for instance \cite{book}), a basis in $\cal K$ is
given by the spin-network states
\begin{eqnarray}\label{base}
\!\!\psi_{{\mathbf j}}
({\mathbf U})\!\!\!&=&\!\!\! \langle {\mathbf U}| 
{\mathbf j} \rangle \\ \nonumber
&=& \!\!\!
R^{j_{12}}_{io}(U_{12}) \,
R^{j_{13}}_{jp}(U_{13})\,
R^{j_{14}}_{kq}(U_{14}) \,
R^{j_{23}}_{lr}(U_{23}) \,
R^{j_{24}}_{ms}(U_{24})\,
R^{j_{34}}_{nt}(U_{34}) \ 
v^{ijk}\; v^{olm}\; v^{prn}\; v^{qst},
\end{eqnarray}
where the $R^{j}_{kl}$ are the matrix elements of the $SU(2)$
representation $j$ and $v^{ikl}$ are the normalized invariant tensors. 
The index structure of equation (\ref{base}) is determined by the
geometry of the tetrahedron.  The function $\psi_{{\mathbf
j}}({\mathbf U})$ is the spin-network function for a spin network
having $T^*$ as graph.  (See \cite{book} for details.)

The left invariant vector field on each group can be identified as the
operator associated to the triad field integrated along the edges of
$T$.  The integral of the $SU(2)$-norm of these gives the length of
the edge; therefore the Casimir operators $C_{pq}$ of the $(pq)$-th
group
\begin{equation}
C_{pq}\ | {\mathbf j} \rangle = j_{pq}(j_{pq}+1)\ | {\mathbf j} \rangle
\end{equation}
can then be naturally identified as the operator giving the length
square of the edge $e^{pq}$ \cite{cpr}.  The tensor structure of the
algebra of the $SU(2)$ representations implements the triangular
relations satisfied by the length.  The spectrum of the length of the
edges $e_{pq}$ is therefore given by
\begin{equation}
l_{pq}=\sqrt{j_{pq}(j_{pq}+1)}.
\end{equation}
The fact that the lengths have discrete spectrum is an immediate
consequence of their conjugate variables being angles, and thus vary
on a compact domain.  Following \cite{book}, we can interpret the
spectral properties of the partial observables as physical predictions
of the quantum model.

\subsection{Dynamics}

The quantum dynamics is completely captured by the propagator
\cite{feynhibbs}.  In a general relativistic theory, the propagator is
formally expressed as the function of the boundary variables given by
(\ref{K}).  Recall, however, that in general the propagator is
\emph{not} a function of classical boundary variables; the reason is
that the boundary quantities may fail to have continuous spectrum.  If
they have discrete spectrum, the propagator depends on the
\emph{quantum numbers} that label the discrete eigenvectors of the
boundary quantities, and not on the corresponding continuous classical
variables \cite{book}.  In our case, the propagator can be written in
the basis (\ref{base}), where it will be a function $W(j_{pq})$.

To find this function, recall that the classical dynamics requires
three-dimensional space to be flat.  This means that any parallel
transport along a three-dimensional closed path must be trivial. 
Consider the four ``elementary" closed paths $\gamma_{p}$ on $T^*$,
where $\gamma_{4}$ is defined by the sequence of edges
$e_{12}e_{23}e_{31}$, that circle the vertex opposite to the face
$f^{4}$, and so on.  The flatness requirement is expressed by the
parallel transport around each of these paths being trivial
\begin{equation}
U_{12}U_{23}U_{31} = \mathbf1,
\end{equation}
and similarly for the other three.  Therefore we can write the 
3d flatness requirement (the Einstein equations) in the form 
\begin{equation}
(U_{pq}U_{qr}U_{rp} - \mathbf1) = 0
\end{equation}
where $p\ne q\ne r$.  We then write the main dynamical equation of the
quantum theory in the form
\begin{equation}
(U_{pq}U_{qr}U_{rp} - \mathbf1)\ \psi_{0}({\mathbf U}) = 
0,
\end{equation}
which can be interpreted as a Wheeler-D toeWitt equation.  Its general
solution is
\begin{equation}
\psi_{0}({\mathbf U})= f({\mathbf U}) \ \prod_{pqr}\delta(U_{pq}U_{qr}U_{rp}),
\end{equation}
where the delta function is the one on the group (for the Haar
measure) and $f({\mathbf U})$ is an arbitrary gauge invariant 
function.  This equation
defines the physical states $\psi_{0}$ that solve the dynamics of the
theory.  To express these states in the ${\mathbf j}$ basis, we simply
project them on the basis states (\ref{base})
\begin{equation}
\psi_{0}({\mathbf j}) = 
\int d{\mathbf U} \ \bar\psi_{{\mathbf j}}({\mathbf U})\ f({\mathbf U}) \
\prod_{pqr}\delta(U_{pq}U_{qr}U_{rp}).
\end{equation}
It is easy to see that by gauge invariance, we can gauge fix
\emph{all} $U_{pq}$ to unity in the integral, giving
\begin{eqnarray}
\psi_{0}({\mathbf j}) &=& \int d{\mathbf U}  \ 
\bar\psi_{{\mathbf j}}({\mathbf U})\ f({\mathbf U}) \ 
\prod_{pq}\delta(U_{pq}) \nonumber\\ &=& c\ \psi_{{\mathbf j}}(1) \nonumber\\
&=& c\ \delta_{io} \delta_{jp} \delta_{kq} \delta_{lr} \delta_{ms}
\delta_{nt}\ v^{ijk}\; v^{olm}\; v^{prn}\; v^{qst} .
\end{eqnarray}
The constant $c=f(\mathbf{{\mathbf 1}})$ can be absorbed in the normalization. 
The last line is the definition of the Wigner 6-$j$ symbol, usually
written as
\begin{equation}
\psi_{0}({\mathbf j}) =\left(\begin{array}{ccc}
    j_{12} & j_{13} & j_{14}  \\
    j_{34} & j_{24} & j_{23} 
\end{array}   \right) \equiv \langle {\mathbf j} | 0 \rangle.
\label{vuoto}
\end{equation}
Thus, we conclude that there is a single state $|0 \rangle$ (up to
normalization) in $\cal K$ that solves the dynamics, and that this
state is proportional to the Wigner 6-$j$ symbol.

The physical amplitude of an arbitrary kinematical state $\psi\in{\cal
H}$ is determined by its projection on the state that solve the
dynamical equation, namely by its projection on the state $|0\rangle$
\begin{equation}
A(\psi)=\langle 0 |\psi\rangle.
\end{equation}
The state $|0\rangle$ is called the ``non-perturbative" vacuum state
\cite{book}.  It expresses the dynamics of the theory.  In other
words, the physical amplitude for having the boundary configuration
$j_{pq}$ is the Wigner 6-$j$ symbol.  Namely the propagator of the
theory is Wigner 6-$j$ symbol
\begin{equation}
W({\mathbf j}) \equiv \langle {\mathbf j} | 0 \rangle = \left(\begin{array}{ccc}
j_{12} & j_{13} & j_{14} \\
    j_{34} & j_{24} & j_{23} 
\end{array}   \right).
\label{W6}
\end{equation}

Now, this is precisely the result obtained by Ponzano and Regge on the
basis of a physical ansatz on the discretization of the lengths, and a
discretization of the Einstein-Hilbert action \cite{Ponzano:1968} ! 
(In our minimalist model, the functional integral (\ref{K}) is
trivial because there are no bulk degrees of freedom.  Its result is
therefore proportional to the exponential of the action.  Ponzano and
Regge found that the Wigner 6-$j$ symbol (\ref{W6}) can in fact be
viewed as a discretization of (the real part) of the exponential of
the action.)  The result is also equivalent (up to a phase) to the
specialization to a single tetrahedron of the boundary amplitude
computed in \cite{ooguri} and in \cite{loug}.  In the present case,
the discretization of the length is not introduced as an ansatz, but
it is a standard quantum-mechanical consequence of the conjugate
variable being an angle.

\subsection{Quantum equilateral tetrahedron}

So far, we have considered an arbitrary quantum tetrahedron.  We now
specialize the formalism to the case of an equilateral tetrahedron. 
The simplest way to do so is to restrict our attention to the states
where four of the six edge lengths are equal.  More precisely, we put
\begin{eqnarray}
j_{a} &\equiv& j_{13}, \nonumber\\
j_{b} &\equiv& j_{24}, \nonumber\\
j_{c} &\equiv& j_{12}=j_{23}=j_{34}=j_{41}
\end{eqnarray} 
and we consider only the states 
\begin{eqnarray}
|j_{a},j_{b},j_{c}\rangle 
=|j_{c},j_{a},j_{c},j_{c},j_{b},j_{c}\rangle.
\end{eqnarray} 
Accordingly, we we restrict the states $\psi(U_{pq})$ to the subset of
$(SU(2))^6$ determined by $U_{12}=U_{23}=U_{34}=U_{41}$.  We write
\begin{eqnarray}
U_{a} &\equiv& U_{13}, \nonumber\\
U_{b} &\equiv& U_{24}, \nonumber\\
U_{c} &\equiv& U_{12}=U_{23}=U_{34}=U_{41}.  
\end{eqnarray}
The gauge transformations that preserve the resulting subspace are the
ones for which
\begin{eqnarray}
V_{1}= V_{3} &\equiv& V_{a}, \nonumber\\
V_{2}=V_{4} &\equiv& V_{b},
\end{eqnarray}
under which the states $\psi(U_{a},U_{b},U_{c})$ transform as
\begin{eqnarray}
\psi(U_{a},U_{b},U_{c}) \to \psi(V_{a}U_{a}V_{a}^{-1},
V_{b}U_{b}V_{b}^{-1},V_{a}U_{c}V_{b}^{-1}). 
\end{eqnarray} 
Using these gauge transformations, we can transform $U_{a}, U_{b},
U_{c}$ to three rotation around three orthogonal axis, of three angles
$k_{a}, k_{b}, k_{c}$.  The interpretation of these angles is simple. 
Since spacetime is flat, we can choose the gauge in which the internal
space is directly identified with flat spacetime.  Then the rotation
along the edge $e_{pq}$ of $T^*$ can be identified as the physical
rotations that one undergoes in crossing the edge $e^{pq}$ of $T$. 
These are precisely the external angles that were denoted $k_{a},
k_{b}, k_{c}$ in the previous section.  For an $SU(2)$ matrix, ${\rm
Tr}(U)=2\cos(\phi/2)$, where $\phi$ is the rotation angle.  Therefore
we can consider the operator
\begin{equation}
T_{a}   \equiv {\rm Tr}(U_{a})=2\cos(k_{a}/2)=2\cos(p_{a}/2),
\end{equation} 
which is now gauge invariant.  The action of this operators is easily
obtained from $SU(2)$ representation theory:
\begin{equation}
T_{a}\ |j_{a},j_{b},j_{c}\rangle
=|j_{a}+1/2,j_{b},j_{c}\rangle
+|j_{a}-1/2,j_{b},j_{c}\rangle,
\end{equation} 
and similarly for the other edges.  In the next section we will show
that the commutator between this operator and the length reproduces
the classical Poisson brackets.

In summary, the boundary Hilbert state $\cal K$ is spanned by the
states $|j_{a},j_{b},j_{c}\rangle$.  The boundary observables $a,b,c,
p_{a}, p_{b}, p_{c}$ that measure the length of the edges of the
tetrahedron and the external angles are represented by Casimir and
trace operators, and the dynamics is given by the propagator
\begin{equation}
W(j_{a},j_{b},j_{c}) =\left(\begin{array}{ccc}
    j_{a} & j_{c} & j_{c}  \\
    j_{b} & j_{c} & j_{c} 
\end{array}   \right),
\label{propagator}
\end{equation}
which expresses the probability amplitude of measuring the lengths
determined by $j_{a}, j_{b}, j_{c}$.  This concludes the definition of
the quantum theory.  The predictions of the theory are given by the
quantization of the lengths and by the relative probability amplitude
(\ref{propagator}).

\section{Time evolution in the quantum theory}

So far, we have viewed our system as a general relativistic system, in
which predictions are expressed in terms of (probabilistic) relations
between boundary partial observables, or probability amplitudes for
boundary configurations.  We now reinterpret the system as a system
evolving in a time variable, as we did in the classical case.  Thus,
we see, say $b$ and $p_{b}$ as initial variables, $a$ and $p_{a}$ as
final variables, and $j_{c}$ at a time parameter. We must identify 
the Hilbert space of the system at fixed time. 

Let us focus on the final state.  This is described by the operators
$C_{a}$ and $T_{a}$ that act on the variable $U_{a}$.  The boundary
Hilbert state $\cal K$, spanned by the states $|j_{a}, j_{b},
j_{c}\rangle$, can be decomposed as (a subspace, because of the
Clebsch--Gordan relations of) the tensor product of three spaces
${\cal K}_{a}, {\cal K}_{b}, {\cal K}_{c}$ spanned by states
$|j_{a}\rangle, |j_{b}\rangle, |j_{c}\rangle$ respectively.  Let us
focus on ${\cal K}_{a}$, which can be interpreted as the state space
at fixed time.

${\cal K}_{a}$ can be simply expressed as the space of the class
functions $\psi(U_{a})$, that is, the functions satisfying
\begin{equation}
\psi(U_{a}) = \psi(V_{a}U_{a}V_{a}^{-1}). 
\end{equation}
The basis $|j_{a}\rangle$ is defined by the characters 
\begin{equation}
\langle U_{a}|j_{a}\rangle = \chi_{j_{a}}(U_{a}) = 
\frac{\sin((j_{a}+1/2)U_{a})}{\sin(U_{a}/2)}. 
\end{equation}
The Casimir and trace operators act as 
\begin{eqnarray}
C_{a}|j_{a}\rangle &=& j_{a}(j_{a}+1)|j_{a}\rangle\\
T_{a}|j_{a}\rangle &=&
|j_{a}+1/2\rangle + |j_{a}-1/2\rangle, 
\end{eqnarray} 
where the second relation is easily derived from the properties of the
characters.  It is convenient to define also the operator
\begin{eqnarray}
S_{a}|j_{a}\rangle &=&
i(|j_{a}+1/2\rangle - |j_{a}-1/2\rangle). 
\end{eqnarray} 
that satisfies $T^2_{a}+S^2_{a}=4$, and is therefore a
function of $T_{a}$
\begin{eqnarray}
S_{a} &=&\sqrt{4-T_{a}^2}.
\end{eqnarray} 
Since we have identified $T_{a}$ with $2\cos(p_{a}/2)$, it follows
that we must identify $S_{a}$ with $2\sin(p_{a}/2)$.  The classical
Poisson brackets
\begin{eqnarray}
\{a,p_{a}\} = 1
\end{eqnarray} 
gives, for $T_{a}=2\cos(p_{a}/2)$,
\begin{eqnarray}
\{a,T_{a}\} = \sin(p_{a}/2) = 1/2 \sqrt{4-T_{a}^2}
\end{eqnarray} 
Consider the operator $J_{a}$ defined by $C_{a}=J_{a}(J_{a}+1)$ and
acting as
\begin{eqnarray}
J_{a}\ |j_{a}\rangle &=& j_{a}\ |j_{a}\rangle.
\end{eqnarray} 
A straightforward calculation gives
\begin{eqnarray}
[J_{a},T_a]  &=& i/2 S_{a}= i/2 \sqrt{4-T_{a}^2}. 
\end{eqnarray} 
Therefore we see that the operators $J_{a}$ and $T_{a}$ define a
linear representation of the classical Poisson algebra defined by the
observables $a$ and $2\cos(p_{a}/2)$.  We have then two options.  The
first is to identify the classical quantity $a$ with the operator
$J_{a}$.  The second is to identify $a$ with the square root of the
Casimir.  Both choices give the correct classical limit, since they
become the same in the limit of large quantum numbers.  The first
gives a quantum theory in which the length is quantized in
half-integers $j_{a}$; the second gives a quantum theory in which the
length is quantized as $\sqrt{j_{a}(j_{a}+1)}$.  We identify the second
choice with the quantization defined in the previous section.

A discrete time evolution is determined by the propagator
(\ref{propagator}), seen as a propagator from the state
$|j_{b}\rangle$ to the state $|j_{a}\rangle$ in a (discrete) time
$j_{c}$.

Recall that in the classical theory the long time evolution drives the
system to the ``Minkowski" configurations where $p_{a}=-\pi$.  Let us
study the quantum evolution for long times.  For $j_c \to \infty$ we
have \cite{Regge}
\begin{equation}
W(j_{a},j_{b},j_{c})=\left(\begin{array}{ccc}
    j_{a} & j_{c} & j_{c}  \\
    j_{b} & j_{c} & j_{c} 
\end{array}   \right) \rightarrow \frac{(-1)^{-(j_a + j_b + 2 
j_c)}}{2 j_c}.
\label{PRInfinito}
\end{equation}
This can be written as 
\begin{equation}
W(j_a,j_b,T)\rightarrow_{T \rightarrow \infty} 
\frac{(-1)^{-(j_a + j_b + 2T)}}{2T}= 
\frac{e^{-2 \pi (iT)}}{2T} \, e^{-i j_b \pi} \, 
e^{-i j_a \pi} = 
\frac{e^{-iE_0 T}}{2T} \, \psi_0(j_{a}) \, \psi_0(j_b).
\label{risultatoOK}                                                 
\end{equation}
That is, for large $T$ evolution projects on the (generalized) state
\begin{equation}
\psi_0(j_{a}) \equiv  \langle j_{a} |0_{M}\rangle = e^{-i j_a \pi} . 
\end{equation}
It is easy to see that this is the generalized eigenstate
of $p_{a}$ with eigenvalue $-\pi$ (since $p_{a}$ itself is not an
operator in the theory, by this we mean, of course, a generalized
eigenstate of $T_{a}=2\cos(p_{a}/2)$ with eigenvalue
$2\cos(-\pi/2)=0$):
\begin{eqnarray}
2\cos(p_{a}/2)\psi_0(j_{a})
&=& T_{a} e^{-i j_a \pi} 
=  e^{-i (j_a+1/2) \pi}+ 
 e^{-i (j_a-1/2) \pi} \nonumber \\
&=& 
e^{-i j_a \pi}(e^{+i\pi/2}+e^{-i\pi/2})=
2\cos(-\pi/2)\psi_0(j_{a}).
\end{eqnarray}
Therefore we have shown that the quantum dynamics converges to the
classical dynamics on long times.  It is appropriate to call
$|0_{M}\rangle$ the ``Minkowski" quantum state, since it minimizes the
energy.

We have shown that the nonperturbative vacuum state $|0\rangle$ in
$\cal K$ become a projector on $|0_{M}\rangle$ in the $T\to\infty$
limit.  We can therefore write the suggestive expression
\begin{equation}
       \lim_{j_{c}\to\infty} |0\rangle = |0_{M}\rangle\langle 0_{M} |.
       \label{formula}
\end{equation}
The bra/ket mismatch is only apparent: the l.h.s is a ket in $\cal K$,
while the r.h.s. is an element of the tensor product between ${\cal
K}_{a}$ and its dual, which can be identified with a subspace of $\cal
K$ under
\begin{equation}
       |j_{a}\rangle\langle j_{b} | \leftrightarrow
       |j_{a},j_{b},j_{c}\rangle.
\end{equation}
See \cite{cdort} and \cite{book} for details.  Equation
(\ref{formula}) is the expression proposed in \cite{cdort} for
computing the Minkowski vacuum state for spinfoam transition
amplitudes.  We see that in the present case this equation is correct. 
Notice, however, that in this euclidean context 
the limit is taken for \emph{real} times.  

Alternatively, we can study the continuous time evolution determined
by quantizing the classical hamiltonian (\ref{hamiltonsPi}).  Notice
that (\ref{hamiltonsPi}) can be easily written in terms of the
operators that we have defined
\begin{eqnarray}
H(a,p_{a},T) & = & 4\pi - 4\arccos\left(\sqrt{C_a}\ 
\frac{1}{\sqrt{4T^2-(4T^2-C_a)\; (T_{a}/2)^2}}\ (T_a/2)\right).
\label{hamiltonsPiq}
\end{eqnarray}
Choosing this ordering (where the inverse and the arccos are defined
by spectral decomposition) we have immediately that the eigenstate of
the $p_{a}$ with eigenvalues $-\pi$ is an eigenstates of the
hamiltonian, with energy $2\pi$, in accord with the corresponding
classical result.  The precise relation between the discrete time
evolution defined by the propagator $W(j_{a},j_{b},j_{c})$ and the
continuum time evolution defined by the Hamiltonian will be studied
elsewhere.

\section{Conclusion}

The model we have considered is obviously extremely simple, and we
cannot derive general conclusions from its analysis.  However, we
think that the structure illustrated by this model does illustrate how
a general covariant quantum field theory can be interpreted. 
Observables can be defined on a closed finite boundary.  The classical
dynamics can be expressed as a set of relations between these
observables.  In the model considered here, these are given in
(\ref{em33}).  The quantum theory can be defined in terms of a
boundary Hilbert space $\cal K$, on which operators representing
boundary observables are defined.  The boundary observables are
partial observables: they represent quantities whose measurements can
be operationally defined in principle, but whose value cannot be
predicted from the knowledge of the state, in general \cite{Partial}. 
The spectral properties of these boundary operators are physical
kinematical predictions of the theory.  Dynamical predictions do not
refer to values of partial observables, but rather to \emph{relations}
between these values.  The quantum dynamics is captured by the
nonperturbative vacuum state $|0 \rangle$, or, equivalently, by the
propagator (\ref{propagator}), which is the expression of $|0 \rangle$
on a basis that diagonalizes boundary observables:
$W(j_{a},j_{b},j_{c})= \langle j_{a},j_{b},j_{c}|0\rangle$.  This
state assigns a probability amplitude to any ensemble of boundary
measurements.

A temporal interpretation of the model is not necessary, but it is
possible \cite{time}.  By interpreting the ``side" length $c$ as a
time variable, the propagator $W(j_{a},j_{b},j_{c})$ can be
interpreted as the transition probability amplitude from the initial
state $|j_{b}\rangle$ to the final state $|j_{a}\rangle$ in a discrete
time $j_{c}$.  The energy that drives this evolution has minimum value
on a state (that we have denote the ``Minkowski" state) that can be
obtained from the propagator: the propagator becomes proportional to a
projector on this state in the large time limit.

The reader might wonder if this structure is tied to the fact that the
theory we have considered here is topological.  We do not think that
this is the case.  The fact that the theory is topological is only at
the origin of the great simplicity that we have found in computing
all features of the model explicitly.  In particular, we have found no
need of computing the functional integral (\ref{K}) explicitly.

The model has some notable specific features.  For instance, energy is
bounded from below as well as a from above.  This fact was first
noticed in 3d quantum gravity by `t Hooft \cite{thooft}.  The
consequence is that the proper time $T$ itself is quantized, as in
loop quantum cosmology \cite{cosmology}.

We think that this simple model illustrates how quantum field theory
can be defined and consistently interpreted in the absence of a
background spacetime.  In particular, the formalism and its
interpretation are well-defined without the need of selecting a time
variable.  The relation with an approximate notion of time evolution
is also illustrated by this model.  The application of these ideas to
full quantum general relativity in four dimensions is of course
nontrivial, but this simple example indicates rather clearly a
direction for defining observables and obtaining predictions in the
general context.

\bigskip
\begin{center}---------\end{center}
\bigskip

Many thanks to Alejandro Perez for important suggestions, and to
Pierre Alain Henry and Bernard Rafaelli for help in numerical
investigations of the behavior of the 6$j$ symbols.  This work is
supported in part by a grant from the Fondazione Angelo Della Riccia.

\end{document}